\documentclass[a4paper,twoside,reqno]{bjp}
\usepackage{graphicx}
\usepackage{cite}
\usepackage{amssymb,amsmath,amscd,amsthm}
\usepackage{times}

\usepackage[bookmarks=false]{hyperref}
\hypersetup{%
    colorlinks=true,        
    linkcolor=blue,          
    citecolor=blue,         
    urlcolor=blue           
    }

\usepackage{geometry}
\allowdisplaybreaks
\usepackage{url}

\def\be{\begin{equation}}
\def\lan{\left\langle}
\def\ran{\right\rangle}
\def\ee{\end{equation}}
\def\barr{\begin{array}}
\def\earr{\end{array}}

\def\l{\left}
\def\r{\right}
\def\dis{\displaystyle}
\def\ed{\end{document}}

\def\f{\frac}
\def\caa{{\cal A}}
\def\chh{{\cal H}}

\def\spin{\frac{1}{2}}

\def\cc{{\cal C}}

\def\chsu4{SO_{sdST}(36) \supset SO_{sST}(6) \oplus SO_{dST}(30)}

\begin{document}
\sloppy

\title{Multiple $SO(5)$ isovector pairing and seniority $Sp(2\Omega)$ multi-$j$ algebras with isospin}

\runningheads{Kota and Sahu}{Multiple pairing algebras with isospin}
 
\begin{start}{%
\author{V.K.B. Kota}{1},
\author{R. Sahu}{2}

\address{Physical Research Laboratory, Ahmedabad 380 009, India}{1}
\address{National Institute of Science and Technology, Palur Hills, Berhampur-761008, Odisha, India}{2}

\received{Day Month Year (Insert date of submission)}
}

\begin{Abstract}

With nucleons occupying several shell model $j$ orbits, the isovector pair creation operator $A^1_\mu$ (creates a two particle state with angular momentum $J=0$ and isospin $T=1$) is no longer unique. Choosing it to be a sum of single-$j$ isovector pair creation operators each with a phase, there will be multiple pair $SO(5)$ algebras with isospin; with $r$ number of $j$ orbits, there will be $2^{r-1}$ $SO(5)$ algebras each with a corresponding complementary $Sp(2\Omega)$ algebra [$2\Omega = \sum_j (2j+1)$] that gives seniority and reduced isospin quantum numbers. Three applications of multiple $SO(5)$ algebras are presented demonstrating the usefulness of considering $SO(5)$ pairing algebras with general sign factors.

\end{Abstract}

\begin{KEY}
pairing, multiple algebras, multi-$j$, $SO(5)$, $Sp(2\Omega)$.
\end{KEY}
\end{start}

\section{Introduction}

With identical nucleons (protons or neutrons) in a single-$j$ shell, the pair creation and annihilation operators ($S_+$ and $S_-$ respectively) and the number operator ($\hat{n}$) generate remarkably the quasi-spin $SU(2)$ algebra. On the other hand, the $Sp(2j+1)$ subalgebra of the Spectrum Generating Algebra (SGA) $U(2j+1)$ is `complementary' to the quasi-spin-$SU(2)$ algebra and the seniority quantum number $v$ that labels the states w.r.t. $Sp(2j+1)$ algebra, i.e. irreducible representation (irreps), corresponds to the quasi-spin quantum number and similarly, particle number $m$ that labels the irreps of $U(2j+1)$ corresponds to the $z$-component of quasi-spin. More importantly, this solves the pairing Hamiltonian $H_p=-S_+S_-$ and allows one to extract $m$ dependence of many particle matrix elements of a given operator; see \cite{Talmi} for full details and applications. 

All the single-$j$ shell results extend to the multi-$j$ shell situation i,e. for identical nucleons occupying several-$j$ orbits, with $(2j+1)$ replaced by $2\Omega=\sum_j (2j+1)$. In this situation, $v$ is called generalized seniority. Now,
a new result is that with $r$ number of $j$-orbits there will be $2^{r-1}$ number of quasi-spin $SU(2)$ and the corresponding $Sp(2\Omega)$ algebras ($S_+=\sum_j \alpha_j S_+(j)$; $\alpha_j=\pm 1$). These multiple quasi-spin algebras (one for each $\alpha_j$ choice) play an important role in deciding selection rules for electric and magnetic multipole operators. See \cite{Kota-bjp1} for details regarding these multiple pairing algebras and \cite{Talmi,akj-1,akj-2,akj-3} for the goodness of multi-$j$ seniority in certain nuclei. Also,  multi-$j$ seniority provides a framework for shell model theory of the interacting boson model \cite{Iac,Tal-Iac}. 

With isospin ($T$) degree of freedom, the algebra changes to the more complex $SO(5)$ algebra.
With $m$-nucleons in a single-$j$ orbit the SGA is
$U(2(2j+1))$ with the 2 coming from isospin. An isospin conserving  subalgebra chain then is $U(2(2j+1)) \supset [U(2j+1) \supset Sp(2j+1)] \otimes SU_T(2)$. Particle number $m$ labels $U(2(2j+1))$, $(m,T)$ label $U(2j+1)$ and $T$ labels $SU_T(2)$ irreps. It is recognized in very early years of shell model that the
isovector pair creation and annihilation operators, isospin
and the number operator generate a $SO(5)$ algebra, which is 'complementary' to the above $Sp(2j+1)$ algebra. The irreps of $SO(5)$ contain two labels and they can be written in terms of the seniority $v$ and reduced isotopic spin $t$ quantum numbers with $(v,t)$ uniquely  labeling the $Sp(2j+1)$ irreps. Another important result is that the isovector pairing Hamiltonian is simply related to the quadratic Casimir invariants of $SO(5)$ and $Sp(2j+1)$. However, an unsatisfactory aspect of the $SO(5)$ algebra of the shell model is that it does not contain isoscalar pair operators in its algebra. For the first papers on single-$j$ shell pairing $Sp(2j+1)$ algebra and the corresponding $SO(5)$ algebra see \cite{Flo-52,Ker-61,Helm-61,Flo-64,He-65a,Pa-65}. Similarly, for technical work on these algebras (for example deriving analytical formulas for the Wigner coefficients of $SO(5)$) see \cite{He-65a,Gi-65,He-65b,HeHe,HeEl,He-89} and for recent applications see \cite{EnVo,Fae-DBD,ibm-32,ibm-33,jpd-112} and references therein. Although many of the single-$j$ shell results extend to the multi-$j$ shell systems, for the multi-$j$ shell situation a crucial aspect is that there will be multiple $SO(5)$ algebras as the isovector pair creation operator here is no longer unique. The purpose of this paper is to introduce and analyze these multiple $SO(5)$ isovector pairing algebras with isospin. 

Before proceeding further, let us add that the general mathematical theory describing complementarity between identical nucleon number non-conserving quasi-spin $SU(2)$ and the number conserving $Sp(2\Omega)$, complementarity between $SO(5)$ and $Sp(2\Omega)$ with isospin and similar complementarity between many other algebras (for example proton-neutron pairing $SO(8)$ algebra with $LST$ coupling and the corresponding number conserving algebras \cite{KoCas,Rowe}) including those for boson systems (see for example \cite{Ko-00,Duke-1}) is due to Neergard based on Howe's general duality theorem \cite{Neegard-1,Neegard-2}; first proof of complementarity is due to Helmers \cite{Helm-61} and later work is due to Rowe et al. \cite{Rowe-2011}. Now we will give a preview.

In Section 2, multiple  pairing $SO(5)$ and $Sp(2\Omega)$ algebras for the multi-$j$ situation are introduced.
Section 3 gives formulas for constructing many-particle matrix elements of the pairing Hamiltonian generating multiple $SO(5)$ algebras. In Section 4 presented are some applications. Finally, Section 5 gives conclusions. 

\section{Multiple multi-$j$ shell $SO(5)$ and $Sp(2\Omega)$ algebras with isospin}

Consider the angular momentum zero coupled isovector pair creation operator $A^1_{\mu}(j)$ for nucleons in a single-$j$ shell, $A^1_{\mu}(j) = \sqrt{(2j+1)/2}\,(a^\dagger_{j\spin} 
a^\dagger_{j\spin}r)^{0,1}_{0,\mu}$\;.
Now, with nucleons in $(j_1,j_2,\ldots,j_r)$ orbits, pair creation operator can be taken as a linear combination of the single-$j$ shell pair creation operators but with different phases giving the generalized isovector pairing operator to be,
\be
\caa^1_{\mu}(\beta) = \dis\sum_{p=1}^r\;\beta_{j_p} A^1_{\mu}(j_p)\;;\;\;\;\{\beta\}=\{\beta_{j_1}, \beta_{j_2}, \ldots, \beta_{j_r}\} = \{\pm 1, \pm 1, \ldots\}\;.
\label{eq.so511}
\ee
Strikingly, the ten operators $\caa^1_\mu(\beta)$, $[\caa^1_{\mu}(\beta)]^\dagger$, $T^1_\mu$ and $Q_0=[\hat{n}-2\Omega]/2$ (equivalently $\hat{n}$) form a pairing $SO^{(\beta)}(5)$ algebra for each $\{\beta\}$ set when $\beta_{j_p} = \pm 1$ as in Eq. (\ref{eq.so511}). Note that $2\Omega=\sum_j (2j+1)$, isospin generators $T^1_\mu = \sum_j \sqrt{(2j+1)/2}\,(a^\dagger_{j \spin} {\tilde{a}}_{j\spin})^{0,1}_{\;\mu}$ and the number operator $\hat{n} = \sum_j \sqrt{2(2j+1)}\,(a^\dagger_{j \spin} {\tilde{a}}_{j\spin})^{0,0}$.
Without loss of generality we can choose $\beta_{j_1}=+1$ and then the remaining $\beta_{j_p}$ will be $\pm 1$. Thus, there
will be $2^{r-1}$ $SO(5)$ algebras. Then, with two $j$ orbits we have two $SO(5)$ algebras $SO^{(+,+)}(5)$ and $SO^{(+,-)}(5)$, with three we have four $SO(5)$ algebras $SO^{(+,+,+)}(5)$, $SO^{(+,+,-)}(5)$, $SO^{(+,-,+)}(5)$ and
$SO^{(+,-,-)}(5)$, with four $j$ orbits there will be eight $SO(5)$ algebras and so on. Significantly,
the isovector pairing Hamiltonians $H_p(\beta) = -G\sum_\mu \caa^1_{\mu}(\beta) \l[\caa^1_{\mu}(\beta)\r]^\dagger$ with
$G$ the pairing strength, is simply related to $\cc_2(SO^{(\beta)}(5))$, the quadratic Casimir invariant of $SO^{(\beta)}(5)$;
$\cc_2(SO^{(\beta)}(5)) = 2 \,\sum_\mu \caa^1_{\mu}(\beta) \l[\caa^1_{\mu}(\beta)\r]^\dagger + T^2 + Q_0(Q_0-3)$.
Further, $SO(5)$ irreps are labeled by  $(\omega_1,\omega_2)$ with $\omega_1$ and $\omega_2$ both integers or half 
integers and $\omega_1 \ge \omega_2 \ge 0$ \cite{Wy-70}. Then, the eigenvalues of $\cc_2(SO^{(\beta)}(5))$ are $\lan \cc_2(SO^{(\beta)}(5))\ran^{(\omega_1,\omega_2)} = \omega_1(\omega_1+3) + \omega_2(\omega_2 +1)$. Expressions for Casimir invariants are given by Racah very early \cite{Rach}. We will now turn to the complementary $Sp^{(\beta)}(2\Omega)$ algebras.  

Consider one-body operators $u^{k,t}_{m_k, m_t}(j_1,j_2)$ defined in terms of the single particle creation and annihilation operators in $(jt)$ space, $u^{k,t}_{m_k, m_t}(j_1,j_2) = (a^\dagger_{j_1\spin} {\tilde{a}}_{j_2\spin})^{k,t}_{m_k, m_t}$ 
where ${\tilde{a}}_{j-m,\spin -m_t} = (-1)^{j-m+\spin-m_t} \,a_{jm,\spin m_t}$. Now, it is easy to prove that the operators $u^{k,t}(j_1,j_2)$ generate the $U(4\Omega)$ SGA. Moreover, we have the subalgebra $U(4\Omega) \supset [U(2\Omega) \supset Sp(2\Omega)] \otimes SU_T(2)$ with $u^{k,0}_{m_{k}, 0}(j_1,j_2)$ operators generating $U(2\Omega)$ and $SU_T(2)$ generating isospin.
Following the results in \cite{Kota-bjp1,Flo-64,He-65b}, it is easy to recognize the generators of $Sp(2\Omega)$ and they are,
\be
\barr{l} 
u^{k,0}_{\mu , 0}(j,j)\;;k=\mbox{odd} \\
V^{k,0}_{\mu ,0}(j_1,j_2) = u^{k,0}_{\mu , 0}(j_1 , j_2) + X(j_1,j_2,k)\;u^{k,0}_{\mu , 0}(j_2,j_1)\;;\;\; j_1 > j_2\;.
\earr \label{eq.so56}
\ee
Now, the most important result is that for every $SO^{(\beta)}(5)$, there will be a complementary $Sp^{(\beta)}(2\Omega)$ algebra with generators given by Eq. (\ref{eq.so56}) provided
\be
\;X(j_1,j_2,k) = (-1)^{j_1+j_2+k}\, \beta_{j_1}\,\beta_{j_2}\;.
\label{eq.so517}
\ee
Proof for the complementarity is given first by Helmers \cite{Helm-61}. Thus, the multiple $SO^{(\beta)}(5)$ algebras that have number non-conserving generators and $Sp^{(\beta)}(2\Omega)$ algebras with only number conserving generators are complementary provided Eq. (\ref{eq.so517}) is satisfied along with Eqs. (\ref{eq.so511}) and (\ref{eq.so56}).

Turning to the irreps,
all $m$ nucleon states transform as the antisymmetric irrep $\{1^m\}$ of $U(4\Omega)$ and the irreps of $U(2\Omega)$ will be two columned irreps $\{2^{m_1} 1^{m_2}\}$ in Young tableaux notation with $2m_1+m_2=m$ and $T=m_2/2$. Similarly, the $Sp(2\Omega)$ irreps are two columned denoted by $\lan 2^{v_1} 1^{v_2}\ran$ giving  $v=2v_1+v_2$ the seniority quantum number and $t=v_2/2$ the reduced isospin. Group theory allows us to obtain $(m,T) \rightarrow (v,t)$ reductions or $(v,t) \rightarrow T$ for a given $m$ \cite{Wy-70}. More importantly, it can be shown 
that $(\omega_1,\omega_2)$ are equivalent to $(v,t)$ giving
$\omega_1 = \Omega-(v/2)$ and $\omega_2=t$.
With all these, the eigenvalues of $H^{(\beta)}_p$ are \cite{Flo-64},
\be
\lan H^{(\beta)}_p\ran^{m,T;v,t} = -\dis\frac{G}{4}\l[(m-v)(2\Omega+3-\frac{m+v}{2}) 
-2T(T+1) +2t(t+1)\r]\;.
\label{eq.so523}
\ee
Note that $SO^{(\beta)}(5) \supset [SO(3) \supset SO(2)]\otimes
U(1)$ with $SO(3)$ generating $T$, $SO(2)$ generating $M_T$ ($T_z$ quantum number (N-Z)/2) and $U(1)$ generating particle number or $H_1=(m-2\Omega)/2$. Then, the eigenstates of $H^{(\beta)}_p$ are 
\be
\l. \l| \Psi_{H^{(\beta)}_p} \r.\ran \Rightarrow \l. \l| (\Omega-\frac{v}{2},t), H_1=\f{m-2\Omega}{2}, T, M_T=\frac{\mbox{N-Z}}{2}
\r.\ran\;.
\label{eq.so524}
\ee
and the labels do not depend on $(\beta)$. However, explicit structure of the wavefunctions do depend on $(\beta)$; see Section 4. Thus, they will effect various selection rules and matrix elements of certain transition operators (see Section 4). Finally, with $SO(5)$ algebra it is possible to factorize $(m,T)$ dependence of various matrix elements \cite{Talmi,He-65b,ibm-32}. 
In order to enumerate the irrep labels in Eq. (\ref{eq.so524}), used is $U(4\Omega) \supset [U(2\Omega) \supset Sp(2\Omega)] \otimes SU_T(2)$ reductions starting with the $\{1^m\}$ irrep of $U(4\Omega)$. All the rules for these are known \cite{Wy-70,Little,KoCas}.
 
\begin{table}[htbp]
\caption[]{Basis states for the $m=6$ system with $T=0$ considered in Section 4. Here, $\Omega_1=6$ and $\Omega_2=5$. See text for details.}\small\smallskip
\begin{tabular}{cc}
\hline
$\;\;\#\;\;$ & $\l.\l|(v_1,t_1)m_1,T_1\;:\;(v_2,t_2)m_2,T_2\;;\; T=0\r.\ran$ \\
\hline
$1$ & $\l.\l|(6,0),6,0\;:\; (0,0)0,0\;;\; 0\r.\ran$ \\
$2$ & $\l.\l|(4,1),6,0\;:\; (0,0)0,0\;;\; 0\r.\ran$ \\
$3$ & $\l.\l|(2,0),6,0\;:\; (0,0)0,0\;;\; 0\r.\ran$ \\
$4$ & $\l.\l|(4,0),4,0\;:\; (2,0)2,0\;;\; 0\r.\ran$ \\
$5$ & $\l.\l|(4,1),4,1\;:\; (0,0)2,1\;;\; 0\r.\ran$ \\
$6$ & $\l.\l|(4,1),4,1\;:\; (2,1)2,1\;;\; 0\r.\ran$ \\
$7$ & $\l.\l|(2,1),4,0\;:\; (2,0)2,0\;;\; 0\r.\ran$ \\
$8$ & $\l.\l|(2,0),4,1\;:\; (0,0)2,1\;;\; 0\r.\ran$ \\
$9$ & $\l.\l|(2,0),4,1\;:\; (2,1)2,1\;;\; 0\r.\ran$ \\
$10$ & $\l.\l|(2,1),4,1\;:\; (0,0)2,1\;;\; 0\r.\ran$ \\
$11$ & $\l.\l|(2,1),4,1\;:\; (2,1)2,1\;;\; 0\r.\ran$ \\
$12$ & $\l.\l|(0,0),4,0\;:\; (2,0)2,0\;;\; 0\r.\ran$ \\
$13$ & $\l.\l|(0,0),2,1\;:\; (4,1)4,1\;;\; 0\r.\ran$ \\
$14$ & $\l.\l|(2,0),2,0\;:\; (4,0)4,0\;;\; 0\r.\ran$ \\
$15$ & $\l.\l|(2,1),2,1\;:\; (4,1)4,1\;;\; 0\r.\ran$ \\
$16$ & $\l.\l|(2,0),2,0\;:\; (0,0)4,0\;;\; 0\r.\ran$ \\
$17$ & $\l.\l|(2,0),2,0\;:\; (2,1)4,0\;;\; 0\r.\ran$ \\
$18$ & $\l.\l|(2,1),2,1\;:\; (2,0)4,1\;;\; 0\r.\ran$ \\
$19$ & $\l.\l|(2,1),2,1\;:\; (2,1)4,1\;;\; 0\r.\ran$ \\
$20$ & $\l.\l|(0,0),2,1\;:\; (0,0)4,1\;;\; 0\r.\ran$ \\
$21$ & $\l.\l|(0,0),2,1\;:\; (2,1)4,1\;;\; 0\r.\ran$ \\
$22$ & $\l.\l|(0,0),0,0\;:\; (6,0)6,0\;;\; 0\r.\ran$ \\
$23$ & $\l.\l|(0,0),0,0\;:\; (4,1)6,0\;;\; 0\r.\ran$ \\
$24$ & $\l.\l|(0,0),0,0\;:\; (2,0)6,0\;;\; 0\r.\ran$ \\
\hline
\end{tabular}
\end{table}

\section{Construction of many-particle matrix for pairing Hamiltonian generating multiple $SO(5)$ algebras}

In order to probe the role of multiple pair $SO^{(\beta)}(5)$ algebras with isospin, we need to obtain the eigenstates of the pairing Hamiltonian $\chh_p$ as a function of $\{\beta\}$'s. 
A convenient basis for constructing the $\chh_p$ matrix is the product basis defined by the single-$j$ shell $SO(5)$ basis. We will illustrate this using two $j$-orbits say $j_1$ and $j_2$. Hereafter, we call the corresponding spaces $a$ and $b$ respectively (or $1$ and $2$). Then, the basis states are,
$\Psi_{ab}(T\;M_T) = \l|(\omega_1^a \omega_2^a)H^a T^a,  
(\omega_1^b \omega_2^b)H^b T^b; T\,M_T\ran$ or equivalently
$\l|(v_1, t_1)m_1 T_1, (v_2, t_2)m_2 T_2; T\,M_T\ran$. 
Given $m$ number of nucleons, with $m_1$ in number in the first orbit and $m_2$ in the second orbit, $m=m_1+m_2$. Note that $\Omega_1=j_1+\f{1}{2}$, $\Omega_2 = j_2+\f{1}{2}$, $H^a=\f{m_1}{2}-\Omega_1$ and $H^b=\f{m_2}{2}-\Omega_2$. Similarly $T^a$ and $T^b$ are the isospins in the two spaces respectively. Now, a general pairing Hamiltonian [with $\caa^1_{\mu}(\alpha) = A^1_\mu(j_1) + \alpha\,A^1_\mu(j_2)]$ is,
\be
\chh_p(\xi,\alpha)= \dis\f{(1-\xi)}{m}\hat{n}_2 - \dis\f{\xi}{m^2} \l\{4\;\dis\sum_\mu \caa^1_{\mu}(\alpha) \l[\caa^1_{\mu}(\alpha)\r]^\dagger\r\}\;.
\label{eq.so530}
\ee
Here, $\hat{n}_2$ is the number operator for the second orbit and
$\xi$ and $\alpha$ are parameters changing from $0$ to $1$ and $+1$ to $-1$ respectively. Note that for $\xi=1$ and $\alpha=+1$ we have a $SO^{(+)}(5)$ algebra in the total two-orbit space and similarly for $\xi=1$ and $\alpha=-1$ the $SO^{(-)}(5)$ algebra. Diagonal matrix elements of $\chh_p$ in our basis follow easily from Eq. (\ref{eq.so523}); note that$\hat{n}_2$ gives $m_2$ and the other part giving diagonal matrix elements is $\sum_\mu [\{A^1_\mu(j_1)[A^1_\mu(j_1)]^\dagger\} + \alpha^2\{A^1_\mu(j_2)[A^1_\mu(j_2)]^\dagger\}]$.
The  off-diagonal matrix elements involve $SO(5) \supset SO(3) \otimes U(1)$ reduced Wigner coefficients and using Eq. (14) of \cite{He-65b} will give,
\be
\barr{l}
\lan (\omega_1^a \omega_2^a)H^a_f T^a_f\;,\;  
(\omega_1^b \omega_2^b)H^b_f T^b_f; T\,M_T \mid \chh_p(\zeta,\alpha)\r. \\
\l. \mid
(\omega_1^a \omega_2^a)H^a_i T^a_i\;,\;  
(\omega_1^b \omega_2^b)H^b_i T^b_i; T\,M_T\ran 
= -\l(\dis\f{4\xi}{m^2}\r)\;(\alpha) \\
\times \l[(\omega_1^a (\omega_1^a+3) + \omega_2^a(\omega_2^a+1)\r]^{1/2}
\;\l[(\omega_1^b (\omega_1^b+3) + \omega_2^b(\omega_2^b+1)\r]^{1/2} \\
\times \; (-1)^{T_f^a+T_f^b+T+1} \;\dis\sqrt{(2T_f^a+1)(2T_f^b+ 1)}\;\l\{\barr{ccc} T & T_f^b & T_f^a \\ 1 & T_i^a & T_i^b \earr \r\} \\
\times \;\;\lan (\omega_1^a \omega_2^a)H^a_i T^a_i\;\;(11)1,1\;\;\mid\mid (\omega_1^a \omega_2^a)H^a_f T^a_f \ran \\
\times \;\;\lan (\omega_1^b \omega_2^b)H^b_i T^b_i\;\;(11)-1,1\;\;\mid\mid (\omega_1^b \omega_2^b)H^b_f T^b_f \ran  
\earr \label{eq.so532}
\ee
for $H_f^a=H_i^a+1$ and $H_f^b=H_i^b-1$; 
$H^a_i=\f{m_1}{2}-\Omega_1$, $H^b_i=\f{m_2}{2}-\Omega_2$. The $\lan ---- \mid\mid ----\ran$ factors above are the $SO(5) \supset SO(3) \otimes U(1)$ reduced Wigner coefficients.
For the $m=6$ system considered in the next Section, the needed Wigner coefficients follow from Tables III in \cite{He-65b} and Table A.1 in \cite{HeHe}. It is important to mention that for simplicity, in Eq. (\ref{eq.so532}) we are not showing the additional label that is required as discussed in \cite{He-65b,HeHe}. This label is called $\beta$ in \cite{HeHe}.
Finally, Eq. (\ref{eq.so532}) can be extended to three or more orbits by using isospin $T$ couplings. Thus, $\chh_p$ construction is possible with multiple $SO^{(\beta})(5)$ algebras provided all the needed Wigner coefficients in Eq. (\ref{eq.so532}) are known. 

\begin{figure}[htb]
	\centerline{\includegraphics[width=4.5in,height=4.5in]{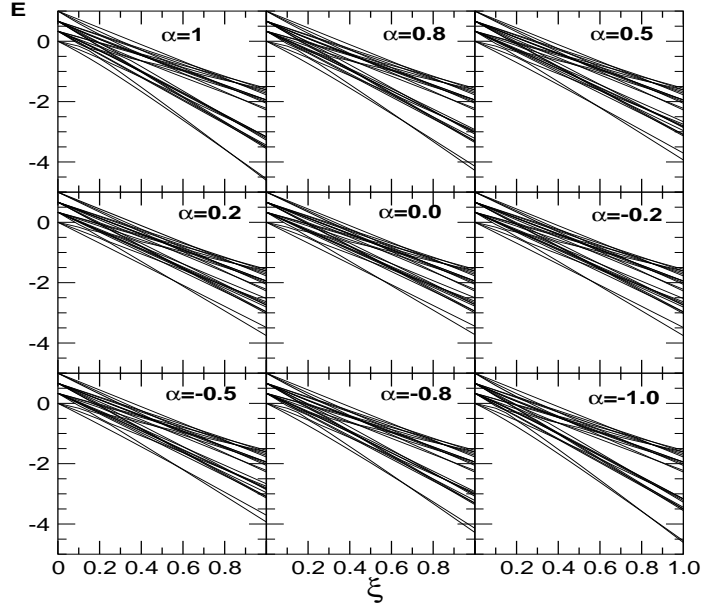}}
\vskip -0.4in
\caption[]{Energy spectra of the pairing Hamiltonian $\chh_p$ in Eq. (\ref{eq.so530}) as a function of $\xi$ and $\alpha$. Note that the energies ($E$) are unitless. See text for details.}
\label{f01}
\end{figure}

\section{Applications of multiple $SO(5)/Sp(2\Omega)$ algebras}

Electromagnetic transition operators $T^{EL}$ and $T^{ML}$ are one-body operators and their $SO(5) \supset [SO_T(3) \supset SO(2)] \otimes U(1)$ tensorial structure is $T^{(\omega_1,\omega_2)}_{H_1,T,M_T}$ with
$(\omega_1 \omega_2) = (11) \oplus(10) \oplus (00)$. More importantly, Eqs. (\ref{eq.so56}) and (\ref{eq.so517}) show that it is possible for $T^{EL}$ and $T^{ML}$ to be  generators of $Sp^{(\beta)}(2\Omega)$ giving selection rules under certain conditions. We have the following results: (i) isovector parts of $T^{EL}$ and $T^{ML}$ will not be $Sp^{(\beta)}(2\Omega)$ scalars as the generators of this algebras are only isoscalar operators;
(ii) the isoscalar part of $T^{ML}$ with $L$ odd (they preserve parity) or even can be $Sp^{(\beta)}(2\Omega)$ scalars provided $\beta_{j(\ell)} = (-1)^{\ell}$ for the $j(\ell)$ orbits; (iii) the $T^{EL}$ with $L$ even or odd will not be generators of any $Sp^{(\beta)}(2\Omega)$ as the $X(j_f,j_i,L)$ (see Eqs. (\ref{eq.so56} and (\ref{eq.so517})) given by the isoscalar part will not lead to a formula for $\beta_{j_i}$ real.
With the phase choice $\beta_{j(\ell)} = (-1)^\ell$, the selection rule from the generators that they will not change $(v,t)$ or $(\omega_1 \omega_2)$ irreps can be used in experimental tests of this phase choice. Besides this, EL and ML transitions can change seniority only by units of 2, i.e.
transition for $v \rightarrow v, v \pm 2$ states are only allowed. In addition, the $(m,T)$ dependence of say quadrupole moments and $B(E2)$'s can be written down using $SO(5)$ algebra. 

In the second application, let us consider a two level system with first level having $\Omega_1=6$ with $-$ve parity and the second level having $\Omega_2=5$ with $+$ve parity. This is appropriate for nuclei in A=56-80 region so that the ($1p_{3/2}$, $0f_{5/2}$, $1p_{1/2}$) orbits with degenerate single particle levels give the $\Omega_1=6$ orbit (we will call it orbit \#1 or $a$) and $0g_{9/2}$ gives the $\Omega_2=5$ orbit (we will call it orbit \#2 or $b$). 
In our numerical calculations we use the system with six nucleons in the above two orbits. Then, the number of $+$ve parity  basis states for $m=6$ and $T=0$ will be 24 as shown in Table 1. Using these basis states, the matrix for $\chh_p$ defined by Eq. (\ref{eq.so530}) is constructed following the formulation in Section 3. Diagonalization of $\chh_p(\xi=1,\alpha=\pm 1)$ will give eigenvalues that must be same as those given by Eq. (\ref{eq.so523}) with $(m=6,T=0)$ and $(v,t)=(6,0)$, $(4,1)$ and $(2,0)$. The eigenvalues are $0$, $-44/m^2$ and $-84/m^2$ respectively with degeneracies $13$, $9$ and $2$ respectively. 
It is easy to see that the wavefunctions are of the form $\l.\l|(v_1,t_1)(v_2,t_2)(v,t)\gamma,m=6,T=0\r.\ran$ where $\gamma$ are additional labels. Therefore, a sum of $\cc_2(SO^{(a)}(5))$ and $\cc_2(SO^{(b)}(5))$ will remove some of the degeneracies in the spectrum without changing the eigenvectors. By adding a term $-(\xi/m^2)[\cc_2(SO^{(a)}(5))+\cc_2(SO^{(b)}(5))]$ to $\chh_p(\xi,\alpha)$ we have calculated the eigenvalues for the $(m=6,T=0)$ system and shown in Fig. 1 are the energies of the 24 states as a function of $\xi$ for nine $\alpha$ values. For example, wavefunctions for the lowest two degenerate states are,
{\scriptsize{
\be
\barr{l}
\l.\l|\Psi^{m=6,T=0}_1\r.\ran = \sqrt{\f{13}{35}} \l.\l|4 (0,0)0; 2(20)0\r.\ran-\alpha\,\sqrt{\f{16}{35}} \l.\l|2 (0,0)1; 2(20)1\r.\ran+\sqrt{\f{6}{35}} \l.\l|0(0,0)0; 6(20)0\r.\ran\;, \\
\l.\l|\Psi^{m=6,T=0}_2\r.\ran = \sqrt{\f{11}{42}} \l.\l|6 (2,0)0; 0(00)0\r.\ran-\alpha\,\sqrt{\f{20}{42}} \l.\l|4 (2,0)1; 2(00)1\r.\ran+\sqrt{\f{11}{42}} \l.\l|2(2,0)0; 4(00)0\r.\ran\;.
\earr \label{eq.so533}
\ee
}}
\noindent Here the notation used is $\l.\l|m_1(v_1,t_1)T_1;m_2(v_2,t_2)T_2\r.\ran$. Eq. (\ref{eq.so533}) shows the role of $\alpha$, i.e. the two $SO(5)$ algebras. As seen from Fig. 1, clearly by changing $(\xi,\alpha)$ it is possible to study order-chaos-order transitions. Detailed analysis of this including all $T$'s will be reported elsewhere.  

Two-particle transfer strengths form the third application. As an example let us consider removal of a isovector pair from the lowest two states [these are $\Psi_1$ and $\Psi_2$ in Eq. (\ref{eq.so533})] of the $(m=6,T=0)$ system generating the states of $(m=4,T=1)$ system. To study the transfer strengths, we have diagonalized $\chh_p(\xi=1,\alpha=\pm 1)$ in $(m=4,T=1)$ space and the basis states here are 14 in number. Then, the eigenstates belong to $(v,t)=(21)$, $(20)$ and $(41)$ irreps in the 4 nucleon space. There are three, two and nine states respectively with these irreps and the corresponding eigenvalues are $-44/m^2$, $-40/m^2$ and $0$ respectively. The transition operator for example can be chosen to be $P=[A^1_{\mu}(j_i)]^\dagger$ or it can be $[\caa^1_{\mu}(\alpha)]^\dagger$ with $\alpha=+1$ or $-1$. These will not change $(v_1t_1)$ and $(v_2t_2)$ of the states. From Eq. (\ref{eq.so533}) it is easy to see that the transfer is allowed to the two states with $(v,t)=(2,0)$ and these are
\be
\barr{l}
\l.\l|\Phi^{m=4,T=1}_1\r.\ran = \sqrt{\f{1}{2}} \l.\l|4 (2,0)1; 0(00)0\r.\ran+\alpha\,\sqrt{\f{1}{2}} \l.\l|2 (2,0)0; 2(00)1\r.\ran\;, \\
\l.\l|\Phi^{m=4,T=1}_2\r.\ran = \sqrt{\f{2}{5}} \l.\l|0(0,0)0; 4(20)1\r.\ran+\alpha\,\sqrt{\f{3}{5}} \l.\l|2 (0,0)1; 2(20)0\r.\ran \;.
\earr \label{eq.so534}
\ee
With the $\alpha$ dependence in both the six and four particle states [see Eqs. (\ref{eq.so533}) and (\ref{eq.so534})], clearly, the two-particle transfer strengths depend on $\alpha$. In practice we need to add a term in the Hamiltonian that mixes the states $\Psi_1$ and $\Psi_2$ and similarly $\Phi_1$ and $\Phi_2$. This is being investigated and explicit formulas for the transfer strengths will be reported elsewhere.
  
\section{Conclusions}

Extending the previous results \cite{Kota-bjp1} on multiple $SU(2)$ pairing algebras for identical nucleons occupying several $j$-orbits, in this paper it is shown that there are multiple $SO(5)$ pairing algebras for nucleons (with isospin) occupying several $j$-orbits. Further, a method to analyze the results, based on the algebra in \cite{He-65b}, due to multiple SO(5) (or the equivalent $Sp(2\Omega)$) algebras is described. Finally, in three applications are briefly discussed. More detailed investigations of multiple $SO(5)$ algebras and their applications will be reported elsewhere. Present work complements the corresponding investigations without isospin in \cite{Kota-bjp1} and on multiple $SU(3)$ algebras in \cite{msu3-1,msu3-2}.

\section*{Acknowledgements} 

R. Sahu is thankful to SERB of DST (Government of India) for financial support.

\ed